\documentclass[%
 reprint,
 amsmath,amssymb,
 aps,
]{revtex4-2}

\usepackage{amsmath,amssymb}
\usepackage{graphicx}
\usepackage{dcolumn}
\usepackage{bm}
\usepackage{comment}
\usepackage{physics}
\newcommand{\ctext}[1]{\raise0.2ex\hbox{\textcircled{\scriptsize{#1}}}}

\begin{document}

\preprint{APS/123-QED}

\title{Control Scheme for Polarization Circulation Speed Meter Using a Dual-Retardation Waveplate}

\author{Yohei Nishino$^{1,2}$}
 \email{yohei.nishino@grad.nao.ac.jp}

\author{Tomotada Akutsu$^{2}$}
\author{Yoichi Aso$^{2,3}$}
\author{Takayuki Tomaru$^{1,2, 3, 4, 5}$}
\affiliation{$^1$Department of Astronomy, University of Tokyo, Bunkyo, Tokyo 113-0033, Japan}
\affiliation{$^2$Gravitational Wave Science Project, National Astronomical Observatory of Japan (NAOJ), Mitaka City, Tokyo
181-8588, Japan}
\affiliation{$^3$The Graduate University for Advanced Studies (SOKENDAI), Mitaka City, Tokyo 181-8588, Japan}
\affiliation{$^4$Accelerator Laboratory, High Energy Accelerator Research Organization (KEK), Tsukuba City, Ibaraki 305-
0801, Japan}
\affiliation{$^5$Institute for Cosmic Ray Research (ICRR), KAGRA Observatory, The University of Tokyo, Kashiwa City,
Chiba 277-8582, Japan}

\date{\today}

\begin{abstract}
In interferometric gravitational wave detectors, quantum radiation pressure noise, which is a back action of the measurement, limits their sensitivity at low frequencies. Speed meters are one of the techniques to reduce the back action noise and improve the sensitivity. Furthermore, a speed meter detector can surpass the standard quantum limit over a wide range of frequencies. The Polarization Circulation Speed Meter (PCSM) is the latest incarnation of the speed meter concept that requires only a modest modification to the conventional interferometer design. However, its control scheme has not been developed. The main difficulty is the length and alignment control of the cavity formed by the polarization circulation mirror and the input test masses, whose round-trip phase shift should be kept at $\pi$. In this article, we propose a new control scheme for the PCSM using a dual-retardation waveplate, called Dual-Retardance Control (DRC). We also compare the shot noise level of the DRC to another simpler scheme by using mirror dithering. Finally, we design the experimental setup to demonstrate the feasibility of the DRC and show the expected results through transfer function measurements.
\end{abstract}

\maketitle
\section{Introduction}
The sensitivity of the gravitational wave (GW) detectors is fundamentally limited by quantum noise. Especially at low frequencies, after the seismic noise and thermal noise are well suppressed and with the use of a high-power laser, it will be limited by \textit{quantum radiation pressure noise}. This low-frequency-limiting noise gives rise to the standard quantum limit (SQL), which is one of the consequences of Heisenberg's uncertainty relation~\cite{Braginsky1992}. The SQL is a fundamental limit that we cannot overcome by conventional methods, and many techniques to beat it, so-called quantum non-demolition (QND) measurement, have been studied~\cite{Braginsky1992, PhysRevD.65.022002}. \\
\indent Speed meters are one of the QND measurements. The concept was first proposed by Braginsky and Khalili~\cite{BRAGINSKY1990251}, and many practical implementations have been investigated~\cite{PhysRevD.61.044002, PhysRevD.67.122004, PhysRevD.66.122004, 2017CQGra..34b4001H, PhysRevD.67.122004, PhysRevD.69.102003, 2017CQGra..34b4001H, https://doi.org/10.48550/arxiv.gr-qc/0211088, PhysRevD.87.096008, SD2018, Knyazev_2018, PhysRevD.86.062001}. The amplitude fluctuations of vacuum fields entering from the anti-symmetric (AS) port of an interferometer are coupled with the pump laser and kick the mirrors randomly, which appears as the quantum radiation pressure noise~\cite{PhysRevD.65.022002}. In speed meters, the vacuum field interacts with the mirror twice with opposite signs. Taking into account the sloshing time $\tau$, which is an interval of two measurements, the back-action force applied on the mirror is~\cite{Danilishin_2012}:
\begin{align}
    \hat{F}_\mathrm{b.a.}(\Omega) \simeq -i\Omega\tau\frac{2\hat{I}_c(\Omega)}{c}
\end{align}
at low frequencies $\Omega \ll 1/\tau$. $\hat{I}_c$ is the fluctuation part of the circulating laser power via the coupling between vacuum fluctuation and the pump laser and $c$ is the speed of light. The signal is proportional to the mean velocity ($\bar{v}$) at low frequencies
\begin{align}
    \phi(t) &\propto \hat{x}(t+\tau)-\hat{x}(t)\\
    &\sim \tau \bar{v},
\end{align}
where $\phi(t)$ is the phase modulation of light and $x(t)$ is the displacement of the mirrors.
Note that the velocity measurement reduces the amount of gravitational wave signal, but in terms of the signal-to-noise ratio, it shows better performance than position measurements at low frequencies.\\
\indent The advantage of speed meters is a broadband-sensitivity improvement at low frequencies. Also by combination with a balanced homodyne detection (BHD), it can go beyond the free mass SQL. It is worth noting that it does not need frequency-dependent homodyne angles, which means we do not need additional filter cavities~\cite{PhysRevD.65.022002}. All the noise-reduction processes happen inside the interferometer, so it is more robust to losses~\cite{Hild2017}.\\
\indent The Polarization Circulation Speed Meter (PCSM) is a new speed meter design proposed by Danilishin \textit{et al}~\cite{SD2018} (see Fig.~\ref{PCSM}). The PCSM only needs a small modification in the AS port; there is no need to modify the central interferometer. Under the current situation that all the large-scale GW detectors are based on the Dual-Recycled Fabry-P\'erot Michelson Interferometer (DRFPMI) and an assumption that it will not be largely changed soon, this design is the most promising candidate for the practical implementation of a speed meter. However, the control scheme has not been investigated yet. \\
\indent There are two main issues to be solved to achieve PCSM. The first issue is that DC component of differential arm signal is reduced to zero in an ideal speed meter, making it hard to control the interferometer. This is a generic problem in speed meters since they reduce the back action noise by deliberately decreasing the mirror motion signal. To obtain the DC signal for the differential arm (DARM) control, one needs to add loss to the interferometer that deteriorates its performance from the ideal case (detailed analysis has been done in ref.~\cite{Danilishin_2015} in the case of the Sagnac-type speed meter). The DARM control for speed meters has already been demonstrated in the proof-of-principle experiment of the Sagnac speed meter in Glasgow group~\cite{Leavey2016}.\\
\indent The second issue is inherent to the PCSM scheme, i.e. we need to keep the round-trip phase shift from the input test masses (ITMs) to the polarization circulation mirror (PCM) at $\pi$ to flip the sign of the second interaction. In this paper, we focus on this issue and propose a new scheme to control the phase shift using a dual-retardation waveplate and an auxiliary laser. These components enable us to obtain a Pound-Drever-Hall (PDH) signal~\cite{1983ApPhB..31...97D} of the cavity length formed by the PCM and ITMs (Polarization Circulation Cavity: PCC). The PDH method is commonly used in the GW detectors and gives us high stability of the PCC length/alignment control. Also, this scheme is well compatible with the balanced homodyne detection (BHD), which is a signal detection scheme planned to be used in the future.\\
\indent The outline of this paper is as follows: in Section~\ref{Background and issues} we show the details of the PCSM and the difficulties of its control. In Section~\ref{DRC}, we propose a new control scheme, DRC, and in Section~\ref{Characterization} we analyze the theoretical performance of the DRC and compare the shot noise levels of the control signals between the DRC and another candidate, the dithering control. In Section~\ref{Experiment}, we show a possible experimental setup for the layout for the demonstration of the DRC. Finally, we give discussions in Section~\ref{discussion} and conclusions in Section~\ref{conclusion}.

\section{Background and issues}\label{Background and issues}
In this section, we review the mechanism of the PCSM, whose detailed study is shown in Danilishin \textit{et al}., 2018~\cite{SD2018}, and explain the inherent difficulties in the PCC control.
\subsection{PCSM}\label{sec_PCSM}
\begin{figure}[h]
\includegraphics[scale=0.7]{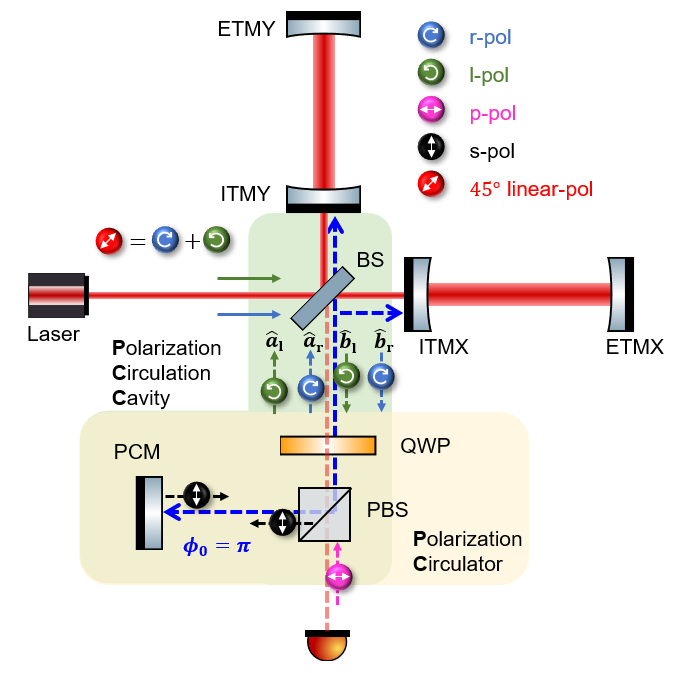}
\caption{\label{PCSM} \textbf{Configuration of the PCSM~\cite{SD2018}.} The QWP converts the polarization state of the vacuum so that it experiences the interferometer twice. The PC is a set of the QWP, PBS, and PCM, and the PCC is a cavity formed by the PCM and the ITMs with the QWP and PBS inside. (E)ITMs stand for (end) input test masses.}
\end{figure}

The schematic design of the PCSM is shown in Fig.~\ref{PCSM}. The main interferometer is the same as the conventional Fabry-Perot Michelson style position meter, but the AS port has two polarization optics, a quarter-wave plate (QWP) and a polarization beam splitter (PBS). These components together with the PCM are collectively called the polarization circulator (PC). The cavity formed by the PCM and the input test masses (ITMs) is called the polarization circulation cavity (PCC). The linear $p$-polarized ($p$-pol) vacuum fluctuation entering from the AS port is converted into the left-polarization vacuum ($l$-pol, denoted by $\hat{a}_l$) by the QWP, couples with the pump laser and kicks the mirror randomly. Then the vacuum field (denoted by $\hat{b}_l$) returns to the AS port and is converted into $s$-polarization ($s$-pol). This $s$-pol beam is reflected by the PBS, thus making a round trip to the PCM. Then it is converted into right-polarization ($r$-pol) by the QWP before coming back to the BS as a field denoted by $a_r$. The $r$-pol beam kicks the mirror again, comes back to the AS port (denoted by $\hat{b}_r$), and finally goes through the PBS. The round-trip phase shift between the ITMs and PCM is kept to $\pi$, so the radiation pressure forces given by $\hat{a}_r$ and $\hat{a}_l$ have opposite signs and cancel out each other. 

\subsection{Difficulties in PCC control}
The PDH method is a commonly used scheme to stabilize the distance between two mirrors can be stabilized on a nanometer scale by the PDH method~\cite{1983ApPhB..31...97D}. All the second-generation GW detectors make full use of this technique to control many degrees of freedom, including the signal recycling cavity (SRC). To control the SRC in the resonant sideband extraction configuration, radio-frequency (RF) sidebands generated by an electro-optic modulator (EOM) are used to sense the length fluctuation of the SRC. \\
\begin{figure*}
\includegraphics[scale=0.8]{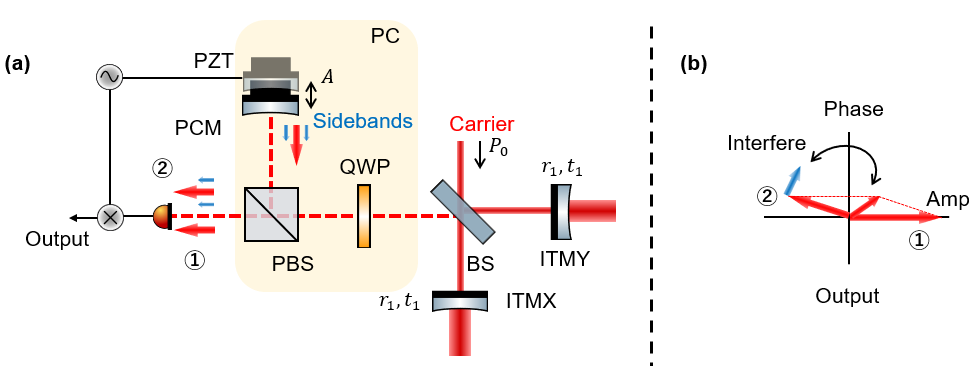}
\caption{\label{Dither} \textbf{Conceptual illustration of the dithering control.} (a) describes the AS port of the PCSM. $r$-pol beam from the interferometer transmits through the PBS as $p$-pol as denoted by \textcircled{\scriptsize 1}. $l$-pol beam is recycled by the PC and transmits through the PBS after the second circulation (see Section \ref{sec_PCSM}). A local oscillator is connected to a PZT behind the PCM to modulate the position of the PCSM. It generates modulation sidebands around the carrier as denoted by \textcircled{\scriptsize 2}. (b) is a phaser diagram for the PBS transmission. A sum of the carrier \textcircled{\scriptsize 1} and \textcircled{\scriptsize 2} has a phase component when the PCM is shifted from the best position. Taking an interference between the carrier and the generated sidebands, one can get an error signal of the PCC length.}
\end{figure*}
\indent However, PCC cannot be locked in the same manner, because the IR beam can circulate inside the PCC twice at most due to the QWP and PBS. It means the finesse of the PCC is almost $0$. This is a serious problem since one cannot effectively amplify signal sidebands. In short, one cannot use the PDH method.\\
\indent As shown above, the PCSM needs a control scheme for the PCC. One simple solution is modulating the PCM position to generate sidebands on the carrier and demodulating the output from the AS port. It is what we call 'dithering' (see Fig.~\ref{Dither}). Detuning the arm cavities and leaking some amount of the DC light (DC offset, see ref.~\cite{DCreadout}), the DC value of the output is zero if the round-trip phase shift in the PCC is kept at $\pi$. Then if the position of the PCM is shifted from the optimal position, non-zero DC signals appear. Taking a beat between the sidebands and the DC offset, one can obtain an error signal.\\
\indent This method is simple but has several problems.
In the first place, mechanical modulation onto the PCM adds noises to the signal sideband, since the modulation frequency of the mirror dithering is $\sim 10$ kHz at most. Secondly, one cannot expect a high signal-to-noise ratio (SNR) in the error signal as shown in Section \ref{Estimation of shot noise level}. The amount of light reaching the AS port is limited, so DC offsets are needed to increase the SNR. However, in future GW detectors, one might not need DC offsets thanks to balanced homodyne detection (BHD), which is also critical for speed meters. To make full use of this advantage, a scheme without DC offsets is preferable. Lastly, it is not sensitive to alignment fluctuations. For these reasons, we propose an alternative scheme for the PCC that can also yield alignment control signals.

\section{Dual-Retardance Control} \label{DRC}
\subsection{Idea}
\begin{figure}[h]
\includegraphics[scale=0.8]{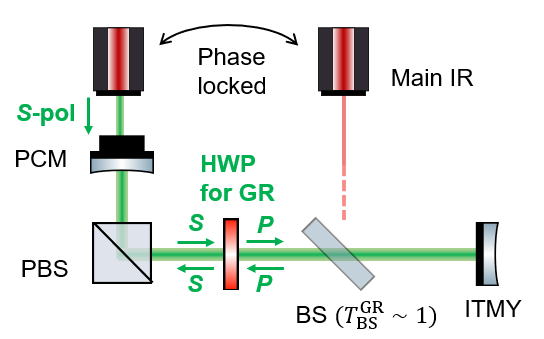}
\caption{\label{PCC} \textbf{Conceptual illustration of the DRC.} We prepare a waveplate that works as a QWP for the carrier and HWP for GR. It changes the polarization from $s$-pol to $p$-pol or $p$-pol to $s$-pol, but it can be kept to only $s$-pol between the PCM and the HWP.}
\end{figure}

 The main obstacle is that the QWP changes the polarization so that the PBS can transmit half of the IR light. If the waveplate does not change the polarization state in one-way transmission or \textit{keep the state so that the PBS does not discard any light}, one can form a cavity with the PCM and ITMs. It can be achieved by a dual-retardance waveplate that works as a QWP for IR but as a HWP for a green (GR) laser.\\
\indent The retardance of a waveplate is described as:
\begin{align}
    \phi_\mathrm{ret} = 2\pi\frac{(n_s-n_f)d}{\lambda_0},
\end{align}
where $n_s$ and $n_f$ are refractive indices along the slow and fast axes respectively, $d$ is the thickness of the waveplate and $\lambda_0$ is the wavelength of the light. In this simple assumption, when it works as a QWP at the wavelength of $\lambda_0$, it should work as a HWP at the wavelength of $\lambda_0/2$. A HWP does not change the polarization state by round-trip transmission (see Fig.~\ref{PCC}). If we inject a laser with half the wavelength of the main interferometer beam from behind the PCM with $s$-pol, it can resonate inside the PCC. Practically, the refractive index has a wavelength dependence, so it is critical to manufacture the dual-retardation waveplate. We call this scheme the dual-retardance control (DRC).\\
\indent The DRC solves the issues discussed in the previous section. With DRC, we can extract a high-SNR error signal for the PCC without using mechanical modulation. We can also obtain alignment signals through wave-front sensing.
\subsection{Setup}
As shown in Fig.~\ref{PCC}, the GR laser frequency ($\omega_\mathrm{GR}$) should be phase-locked to the main IR frequency ($\omega_0$), with a tunable frequency offset of $\omega_\mathrm{off}$:
\begin{align}
    \omega_\mathrm{GR} = 2(\omega_0+\omega_\mathrm{off}).
\end{align}
 Also, the IR and GR beams have to be co-aligned. Then one needs an additional cavity outside the PCC to make them share the same beam path. For example, there is a ring cavity to co-align the IR and GR beam paths in Fig.~\ref{experimental_setup}. The arm cavity can also be used for the path-sharing process. The transmissivity of the BS for GR is set to $\sim 1$ for simplicity. \\
\indent Even though the paths seem to completely overlap, the optical path length of the PCC for the IR ($l_\mathrm{PCC}^\mathrm{IR}$) may not exactly be the same as that for the GR ($l_\mathrm{PCC}^\mathrm{GR}$) due to the dispersion of materials:
\begin{align}
    l_\mathrm{PCC}^\mathrm{GR} = l_\mathrm{PCC}^\mathrm{IR} + \delta l_\mathrm{PCC},
\end{align}
where $\delta l_\mathrm{PCC}$ the difference of the optical path lengths.
Adding a frequency offset $\omega_\mathrm{off}$, the round-trip phase shift in the PCC for the GR is: 
\begin{align}
    \phi_\mathrm{GR} &= 2\omega_\mathrm{GR}l_\mathrm{PCC}^\mathrm{GR}/c\\
    &= 2\left[2(\omega_0+\omega_\mathrm{off})+\delta\omega\right](l_\mathrm{PCC}^\mathrm{IR} + \delta l_\mathrm{PCC})/c\\
    &= \frac{4\omega_0 l_\mathrm{PCC}^\mathrm{IR}}{c}+\frac{4\omega_0 \delta l_\mathrm{PCC}}{c} + \frac{4\omega_\mathrm{off}l_\mathrm{PCC}^\mathrm{GR}}{c}+\frac{2\delta\omega l_\mathrm{PCC}^\mathrm{GR}}{c} \label{eq:phi_GR}
\end{align}
The first term is a phase shift if there is no dispersion. The second term is a shift due to the dispersion, and the third term is a phase compensation by the frequency offset. The fourth term is a phase noise in the Phase Locked Loop (PLL), which results in the average PCC fluctuation (see $\epsilon$ in Eq. (7) in ref.~\cite{SD2018}).
\subsection{Lock acquisition}
\begin{figure}
    \centering
    \includegraphics[scale=0.7]{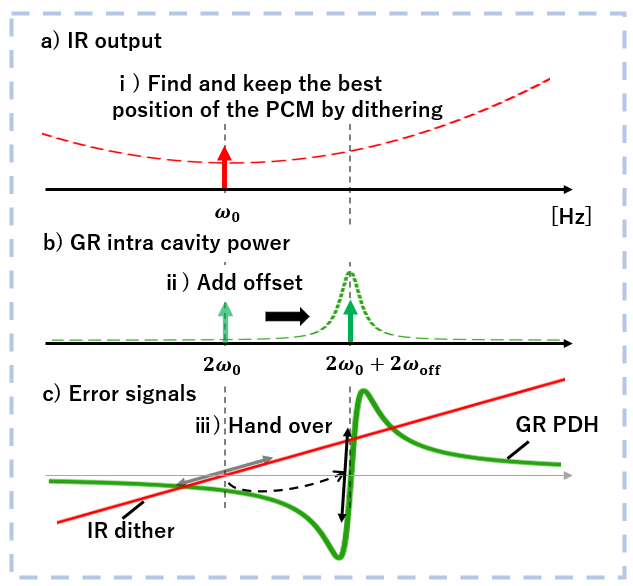}
    \caption{\label{Handover}\textbf{Toy picture of the lock acquisition.} a) The DC output of the PBS transmission. b) GR intra-cavity power. c) The solid red line is the dithering signal by dithering and the solid green line is the GR PDH signal. After adding offsets, one can hand over the error signals to the GR PDH which is steeper than the dithering signal.}
\end{figure}

In order to draw the PCC into the operational condition ($\phi_0=\pi$), we need to follow a set of certain steps. We call the procedure "lock acquisition". \\
\indent The conceptual figure of the lock acquisition is shown in Fig.~\ref{Handover}.
First, using the dithering method, the round-trip phase shift of the IR $\phi_\mathrm{IR}$ is locked to $\pi$ (see the denotation (i) in Fig.~\ref{Handover}). This is realized when the first term in Eq.~(\ref{eq:phi_GR}) satisfies the following condition:
\begin{align}
    2\phi_\mathrm{IR} = \frac{4\omega_0 l_\mathrm{PCC}^\mathrm{IR}}{c} \equiv 0\ (\mathrm{mod}\ 2\pi).
\end{align}
One needs to detune the arm cavities to obtain enough DC signals if necessary. The dithering signal is fed-back to the mechanical actuator on the PCM (a PZT, for example).\\
\indent Second, the GR beam is kept at the resonance of the PCC by adding the offset frequency (see the denotation (i\hspace{-.1em}i) in Fig.~\ref{Handover}). It corresponds to the round trip phase shift for the GR in the PCC satisfying below:
\begin{align}
    \phi_\mathrm{GR} \equiv 0\ (\mathrm{mod}\ 2\pi).
\end{align}
The GR PDH signal is fed-back to the frequency actuator on the GR (an acousto-optic modulator, for example).\\
\indent Lastly, one can hand over the error signals to the GR PDH which is steeper than the dithering signal (see the denotation (i\hspace{-.1em}i\hspace{-.1em}i) in Fig.~\ref{Handover}). Given the absolute frequency of the main IR ($\omega_0$) and the optical path difference ($\delta l_\mathrm{PCC}$) is stable enough, the round-trip phase fluctuations for the GR are proportional to the length fluctuation of the PCC. In this final stage, the GR PDH is fed-back to the PCM. Note that the last term in Eq. (\ref{eq:phi_GR}):
\begin{align}
    \delta\phi_\mathrm{PCC} = \frac{2\delta\omega l_\mathrm{PCC}^\mathrm{GR}}{c}
\end{align}
contributes to the noise of the PCC length. After the handover, the dithering and DC offset can be removed.

\section{Theoretical performance} \label{Characterization}
\subsection{Error signal}\label{sec.error_signal}
\begin{figure}[h]
    \centering
    \includegraphics[scale=0.9]{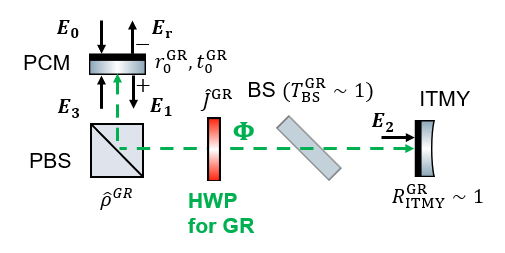}
    \caption{\label{boundary_condition}\textbf{PCC as seen by the green field.} The HWP is represented in the Jones matrix, $\hat{J}^\mathrm{GR}$. The reflectivity of the PBS is also represented in the reflectivity matrix, $\hat{\rho}^\mathrm{GR}$. We assume the BS is transparent for the GR for simplicity}
\end{figure}
\begin{figure}[h]
\includegraphics[scale=0.45]{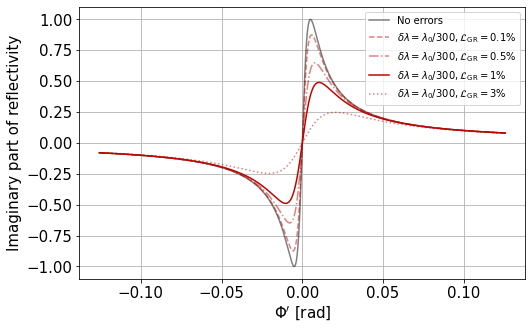}
\caption{\label{DRC_error_various_P} \textbf{Imaginary part of the PCC reflectivity.} Red curves show the imaginary part of  Eq. \ref{eq:r_ss} with various round-trip losses with retardation error of $\lambda_0/300$. The black curve is an error signal without any retardation error and loss.}
\end{figure}

In this section, we analyze the electric fields of a cavity with an HWP and PBS inside to derive the GR PDH signal. We define bases of $p$- and $s$-pol electric fields as:
\begin{align}
    \vb{e}_p = \begin{pmatrix}
        1\\
        0
    \end{pmatrix},\ 
    \vb{e}_s = \begin{pmatrix}
        0 \\
        1
    \end{pmatrix}.
\end{align}
Symbols used in the analysis are shown in Fig.~\ref{boundary_condition}. The reflectivity matrix of the PBS is:
\begin{align}
    \hat{\rho}^\mathrm{GR} = 
    \begin{pmatrix}
    \sqrt{R_p^\mathrm{GR}} & 0 \\
    0 & \sqrt{R_s^\mathrm{GR}}
    \end{pmatrix} 
\end{align}
where $R_s^\mathrm{GR}$ and $R_p^\mathrm{GR}$ is the power reflectivity for $s$-pol and $p$-pol of the PBS. $r_0^\mathrm{GR}, t_0^\mathrm{GR}$ are the amplitude reflectivity and transmissivity of the PCM. The Jones matrix for the $45^\circ$ rotated HWP can be written as:
\begin{align}
    \hat{J}^\mathrm{GR} = \frac{1}{2}
    \begin{pmatrix}
    1+e^{-2i\delta\phi} & 1-e^{-2i\delta\phi} \\
    1-e^{-2i\delta\phi} & 1+e^{-2i\delta\phi}
    \end{pmatrix},
\end{align}
where $\delta\phi$ is the retardation error.\\
Relations between the fields are written as:
\begin{align}
    \vb*{E}_1 &= t_0^\mathrm{GR}\vb*{E}_0+r_0^\mathrm{GR}\vb*{E}_3, \label{8.eq:1}\\
    \vb*{E}_2 &= e^{i\Phi/2}\hat{J}^\mathrm{GR}\hat{\rho}^\mathrm{GR}\vb*{E}_1, \label{8.eq:2}\\
    \vb*{E}_3 &= e^{i\Phi/2}\hat{\rho}^\mathrm{GR}\hat{J}^\mathrm{GR}\vb*{E}_2, \label{8.eq:3}\\
    \vb*{E}_r &= -r_0^\mathrm{GR}\vb*{E}_0+t_0^\mathrm{GR}\vb*{E}_3, \label{8.eq:4}
\end{align}
where $\Phi$ is the round-trip phase shift in the PCC.
Solving those equations, the reflectivity for the $s$-pol at the anti-reflection side of the PCM is:
\begin{align}
    &r_{s\rightarrow s}(\Phi^{'}) = \frac{E_{0, s}}{E_{\mathrm{r}, s}} \\
    &= -r_0^\mathrm{GR}+\frac{(t_0^\mathrm{GR})^2(R_s^\mathrm{GR}\cos\delta\phi e^{i\Phi^{'}}-r_0^\mathrm{GR }R_p^\mathrm{GR} R_s^\mathrm{GR} e^{2i\Phi^{'}})}{\det M} \label{eq:r_ss}
\end{align}
where
\begin{align}
    \Phi^{'}=\Phi-\delta\phi, 
\end{align}
and
\begin{align}
    \det M &= 1-r_0^\mathrm{GR}(R_s^\mathrm{GR}+R_p^\mathrm{GR})\cos\delta\phi e^{i\Phi^{'}} \notag \\
    &\qquad \qquad +(r_0^\mathrm{GR})^2 R_s^\mathrm{GR} R_p^\mathrm{GR} e^{2i\Phi^{'}}.
\end{align}
Here we assumed the reflectivity of the ITMY and the transmissivity of the BS are $\sim 1$ for simplicity. 
While optical components inside the PCC may have their own losses, here we assumed that all the loss is concentrated in the PCM (denoted $\mathcal{L}_\mathrm{GR}$ in Fig.~\ref{DRC_error_various_P}). We set $R_p^\mathrm{GR}$ to 0, which means $p$-pol generated by the retardation error is discarded from the PBS. The imperfection of the $s-$pol reflectivity is counted as a loss on the PCM:
\begin{align}
    \mathcal{L}_s^\mathrm{GR}=1-R_s^\mathrm{GR}.
\end{align}
We show the imaginary part of Eq. (\ref{eq:r_ss}) in Fig.~\ref{DRC_error_various_P} with various round trip losses, which decrease the slope of the error signals.\\
\subsection{Estimation of the shot noise level of the DRC}\label{Estimation of shot noise level}
We compare the shot noise levels of two methods, the dithering control and DRC. The detailed analysis is shown in Appendix~\ref{shotnoise_derivation}. Using Eq.~(\ref{eq:ratio}) and choosing realistic parameters (see Table~\ref{tab:IR}), the ratio of the shot noises of the two methods becomes:
\begin{align}
    \frac{S_L^{\mathrm{Dither}}}{S_L^\mathrm{DRC}} \sim 7.4 \times 10^4.
\end{align}
This large ratio comes from the advantages of using a cavity: the amplification of the phase change by a factor of the finesse of the cavity and the amount of the local oscillator power that can be used for control. 

\section{Experimental demonstration of DRC} \label{Experiment}
\begin{figure*}
\includegraphics[scale=1]{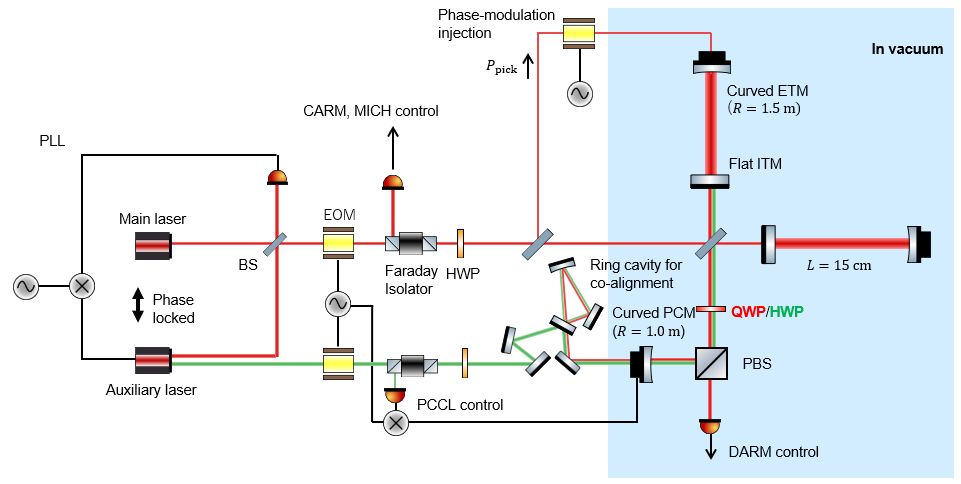}
\caption{\label{experimental_setup} \textbf{Design of an experimental setup for demonstrating the DRC.} The basic configuration is a FPMI with 15 cm arm cavities. The GR laser is phase-locked to the main IR laser and injected from the AR side of the PCM.}
\end{figure*}
\begin{figure}[h]
\includegraphics[scale=0.45]{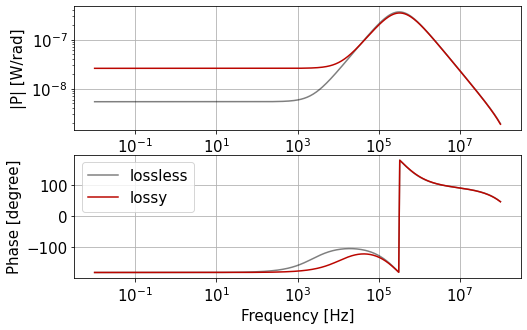}
\caption{\label{Transfer_function} \textbf{Expected transfer functions for the DARM noise injection.} The red curve shows the transfer function when the PCC is assumed to have a loss. The gray curve shows the lossless case.}
\end{figure}

An experimental setup to demonstrate the DRC is shown in Fig.~\ref{experimental_setup}. Possible parameters for IR and GR optics are shown in Table~\ref{tab:IR} and~\ref{tab:GR}, respectively. We aim to confirm the DRC works. The GR laser is generated by the second harmonic generation and phase-locked to the main IR laser. The basic design is a Fabry-P\'erot Michelson Interferometer (FPMI) with $15$-cm-long rigid arm cavities with flat ITMs and curved ETMs. The FPMI part is controlled by the pre-modulation method as used in all the current GW detectors. The error signal of the PCC obtained by the GR PDH is fed back to the PCM. The GR laser frequency is tunable by changing the frequency offset in the PLL.\\
\indent A small fraction of the main IR is picked off after the EOM and injected from the AR side of the ETMY. This light gets phase-modulated through an EOM to generate sidebands that play the role of pseudo-GW signals. \\
\indent The expected transfer function from the phase modulation to the DARM output is shown in Fig.~\ref{Transfer_function}. Given the carrier is resonant in the arm cavities, the amplitude reflectivity of a single arm cavity can be written as:
\begin{align}
    r(\Omega) &\simeq \frac{\gamma_1-\gamma_2+i\Omega}{\gamma_1+\gamma_2-i\Omega},
\end{align}
where $\Omega$ is the frequency of an audio sideband. $\gamma_1$ and $\gamma_2$ are defined as:
\begin{align}
    \gamma_1 &\equiv \frac{cT_\mathrm{ITM}}{4l_\mathrm{arm}}\ (\mathrm{cavity\ pole}),\\
    \gamma_2 &\equiv \frac{c\mathcal{L}_\mathrm{arm}}{4l_\mathrm{arm}}.
\end{align}
$T_\mathrm{ITM}$ is the power transmissivity of the input mirror. $\mathcal{L}_\mathrm{arm}$ is the round-trip power loss of the arm cavity and $l_\mathrm{arm}$ is the length of the arm cavity. Denoting the round-trip power loss in the PCC as $\mathcal{L}_\mathrm{PCC}^{'}$, the transfer function is proportional to:
\begin{align}
    \mathrm{(Output)} &\propto \frac{1- \cos\delta\phi_{\mathrm{PCC}}(1-\mathcal{L}_\mathrm{PCC}^{'})r(\Omega)}{2} \notag\\
    &\simeq \frac{\gamma_2+\mathcal{L}_\mathrm{PCC}\gamma_1/2-i\Omega}{\gamma_1-i\Omega}, \label{eq:29}
\end{align}
where $\mathcal{L}_\mathrm{PCC}$ is the effective-total PCC loss including the round-trip phase fluctuation in the PCC $\delta\phi_{\mathrm{PCC}}$:
\begin{align}
    \mathcal{L}_\mathrm{PCC} &\simeq \mathcal{L}_\mathrm{PCC} ^{'}+\frac{(\delta\phi_{\mathrm{PCC}})^2}{2} \label{eq:30}\\
    &= 2 (\mathcal{L}_\mathrm{BS}+\mathcal{L}_\mathrm{QWP}+\mathcal{L}_\mathrm{PBS}+T_\mathrm{SPBS}+R_\mathrm{PPBS}) \notag\\
    &\qquad\qquad +\mathcal{L}_\mathrm{PCM}+\mathcal{L}_\mathrm{align}+\mathcal{L}_\mathrm{mis} +\frac{(\delta\phi_{\mathrm{PCC}})^2}{2}. \label{eq:31}
\end{align}
In Eq. (\ref{eq:30}), we assume that $\delta\phi_{\mathrm{PCC}}$ is so small that its cosine is approximated as:
\begin{align}
\cos(\delta\phi_\mathrm{PCC})\simeq 1-\frac{(\delta\phi_\mathrm{PCC})^2}{2}.
\end{align}
Definitions of each term in Eq.~(\ref{eq:31}) are shown in Table \ref{tab:loss}. Note that losses in the PCC optical path are doubled due to the polarization circulation, except for the PCM. The mode-mismatching due to the PCM misalignment and the Schnupp asymmetry is also counted as losses. The final term $\delta\phi_\mathrm{PCC}$ is the length fluctuation of the PCC.\\ 
\indent Eq. (\ref{eq:29}) means the PCC loss generates a zero at:
\begin{align}
    \gamma_\mathrm{cut} &= \gamma_2+\frac{\mathcal{L}_\mathrm{PCC}\gamma_1}{2} \\
    &= \frac{c}{4l_\mathrm{arm}}\left[\mathcal{L}_\mathrm{arm}+\frac{\pi\mathcal{L}_\mathrm{PCC}}{\mathcal{F}}\right],
\end{align}
where $\mathcal{F}$ is the finesse of the arm cavity:
\begin{align}
    \mathcal{F} \equiv \frac{2\pi}{T_\mathrm{ITM}}.
\end{align}
In Fig.~\ref{Transfer_function}, we show transfer functions of both lossless and loss-included cases. The cutoffs at low frequencies are generated by the losses. The $\propto 1/f$ structure above the cavity pole is due to the first-order low-pass nature of the arm cavities. Note that even in the lossless case (gray line), we still see a cutoff. It is caused by the transmission of the ETM that is necessary to inject the artificially phase-modulated light.\\
The $\propto f$ structure in the transfer function will be observed if we realize the proposed experiment with the designed parameters. Through the measurement, we can evaluate the performance of the DRC.
\section{Discussion}\label{discussion}
One of the potential issues is the long-term stability of the dispersion of the QWP. It might change due to the heat effect of the laser or environmental temperature fluctuation. Also, beam jitters might also be a source of the noise. It is necessary to test the stability of the PCC control and check how frequently the dithering control needs to be used to ensure the round trip phase of the PCC to be $\pi$.\\
\indent From the perspective of practical implementation, the DRC might conflict with the lock acquisition scheme of the existing detectors. KAGRA, for example, injects auxiliary GR lasers from the center part of the interferometer. To avoid the GR leaking and resonating inside the arm cavity, the DRC sets the ITM transmissivity for the GR as small as possible. Hence it is necessary to find a compromise between them.
\section{Conclusion} \label{conclusion}
In this article, we propose a feasible control scheme for the PCSM using a dual-retardation waveplate and auxiliary laser. We name it DRC. The DRC makes it possible to control the PCC length and alignment. In the DRC, we can get error signals with a higher SNR than the dithering control. Also, the DRC is compatible with the BHD because we do not need the DC offset anymore after the full PCC lock is achieved. After the experimental demonstration of the DRC with rigid cavities, we will proceed to the fully-suspended systems to realize the PCSM in the future GW detectors such as the Einstein Telescope~\cite{Hild_2010}.

\begin{acknowledgments}
We thank Stefan Danilishin, Marc Eisenmann, Kentaro Komori, and Kentaro Somiya for fruitful
discussions. 
\end{acknowledgments}

\begin{table}[b]
\caption{\label{tab:IR}
\textbf{Parameters for IR used for the design of the experiment.}}
\begin{ruledtabular}
\begin{tabular}{ccc}
Parameters & value & Note  \\ \hline
    $\lambda_0$[nm] & 1064  & Nd:YAG \\ 
    $P_0$ [mW] & 50 & IR laser intensity \\ 
    $P_\mathrm{pick}$ [$\mu$W] & 125 & Pick-off laser intensity\footnotemark[1]\\
    $T_\mathrm{ITM}$ & 0.004 & ITM transmissivity\footnotemark[2]\\
    $T_\mathrm{ETM}$ & 30 ppm & ETM transmissivity \\ 
    $T_\mathrm{PCM}$ & $1\ \%$ & PCM transmissivity\\ 
    $R_\mathrm{ITM}$ [m] & $\infty$ & ITM radius of curvature \\ 
    $R_\mathrm{ETM}$ [m] & 1.5  & ETM radius of curvature \\ 
    $R_\mathrm{PCM}$ [m] & 1 & PCM radius of curvature \\ 
    $l_\mathrm{arm}$ [m] & 0.15  & Arm cavity length \\ 
    $l_\mathrm{michx}$ [m] & 0.075 & Length from the BS to ITMX \\
    $l_\mathrm{michy}$ [m] & 0.125 & Length from the BS to ITMY \\
    $l_\mathrm{PCC}$ [m] & 0.307 & Mean PCC length \\ 
    $f_\mathrm{m}$ [MHz] & 47.5 & RF sideband frequency\footnotemark[3]\\ 
    $A$ [nm] & 0.1  & Modulation amplitude \footnotemark[4]\\
    $P_c$ [$\mathrm{\mu}$W] & 100 & DC offset power at the AS port\footnotemark[4] \\ \hline
    $\mathcal{F}$ & $\sim1500$ & Finesse \\
    $f_c$ & $3.2\times10^5$ [Hz] & Cavity pole \\
    $f_\mathrm{cut}$ & $1.7\times10^4$ [Hz] & Cutoff frequency
\end{tabular}
\footnotetext[1]{for phase-noise injection}
\footnotetext[2]{Fused Silica substrate}
\footnotetext[3]{used for the FPMI control}
\footnotetext[4]{see Appendix \ref{shotnoise_derivation}}
\end{ruledtabular}
\end{table}

\begin{table}[t]
\caption{\label{tab:GR}
\textbf{Parameters for GR.}}
\begin{ruledtabular}
\begin{tabular}{ccc}
Parameters & value & Note  \\ \hline
    $\lambda_\mathrm{GR}$ [nm] & 532 & Wavelength of the GR laser \\
    $P_\mathrm{GR}$ [mW] & 20 & GR laser intensity \\ 
    $T_\mathrm{PCM}$ & $0.01$ & PCM transmissivity\\ 
    $T_\mathrm{ITM}$ [m] & $< 10$ ppm & ITM transmissivity \\ 
    $l_\mathrm{PCC}$ [m] & 0.332 & Length from the PCM to ITMY \\ 
    $\beta_\mathrm{m}$ & 0.2 &  Modulation index\footnotemark[1] \\
    $\delta\phi^\mathrm{GR}_\mathrm{ret}$ & $2\pi\lambda_\mathrm{GR}/300$ & QWP retardation error for GR \\ 
    $\mathcal{L}_\mathrm{GR}$ & 3 \% & Total losses in the PCC\\
    $\mathcal{F}_\mathrm{GR}$ & 150 & Finesse of the PCC \\
\end{tabular}
\footnotetext[1]{see Appendix \ref{shotnoise_derivation}}
\end{ruledtabular}
\end{table}
\appendix
\section{Shot noise estimation} \label{shotnoise_derivation}
In the case of the PDH method, the reflection power before demodulation can be written as~\cite{2001AmJPh..69...79B}:
\begin{align}
    P = P_\mathrm{DC} + D\delta l \sin\omega_\mathrm{m}t + \mathcal{O}(2\omega_\mathrm{m}).
\end{align}
$\delta l$ is the length fluctuation of the cavity we want to control and $\omega_\mathrm{m}$ is the frequency of sidebands generated by an EOM. $D$ [W/m] corresponds to the slope of the error signal, which is proportional to the carrier and sideband power $P_\mathrm{c}, P_\mathrm{s}$ and the imaginary part of the reflectivity $\mathrm{Im}[r(\omega)]$:
\begin{align}
    D \propto \sqrt{P_\mathrm{c}P_\mathrm{s}}\mathrm{Im}[r(\omega)].
\end{align}
$P_\mathrm{DC}$ is the DC power, which is the source of the shot noise. The shot noise can be written in the single-sided form as:
\begin{align}
    S_\mathrm{shot} = \sqrt{2e\frac{e\eta P_\mathrm{DC}}{\hbar\omega_0}}\ \mathrm{[A/\sqrt{Hz}]},
\end{align}
where $e$ is the elementary charge and $\eta$ is the quantum efficiency of the photo detector [A/W].
Hence the shot-noise-equivalent length noise is:
\begin{align}
    S_L = \frac{S_\mathrm{shot}}{D}. \label{eq:length_shot_noise}
\end{align}
\indent In the case of the DRC, one can calculate its shot noise level in the same manner. The imaginary part of the reflectivity $r_{s\rightarrow s}$ can be expressed around $\Phi^{'}$ as:
\begin{align}
    \left.\mathrm{Im[r_{s\rightarrow s}(\Phi^{'})]}\right|_{\Phi^\prime=0} &\simeq \left.\frac{\mathrm{d}\mathrm{Im}[r_{s\rightarrow s}(\Phi^\prime)]}{\mathrm{d}\Phi^\prime}\right|_{\Phi^\prime=0}\times \delta\Phi^{'} \\
    &= \left.\frac{\mathrm{d}\mathrm{Im}[r_{s\rightarrow s}(\Phi^\prime)]}{\mathrm{d}\Phi^\prime}\right|_{\Phi^\prime=0}\times \frac{4\omega_0 \delta l_\mathrm{PCC}}{c}.
\end{align}
The slope amplitude $D^\mathrm{DRC}$ can be written as:
\begin{align}
    D^\mathrm{DRC}\delta l_\mathrm{PCC} &= 4\sqrt{P_c P_s} \left.\mathrm{Im[r_{s\rightarrow s}(\Phi^{'})]}\right|_{\Phi^\prime=0} \notag \\
    & = \frac{8\beta_\mathrm{m}\omega_0P_\mathrm{GR}}{c} \left.\frac{\mathrm{d}\mathrm{Im}[r_{s\rightarrow s}(\Phi^\prime)]}{\mathrm{d}\Phi^\prime}\right|_{\Phi^\prime=0}\delta l_\mathrm{PCC} \notag \\
    &= \frac{8\beta_\mathrm{m}\omega_0P_\mathrm{GR}\xi}{c}\delta l_\mathrm{PCC}, \label{eq:D_DRC}
\end{align}
where
\begin{align}
    \xi &\equiv \left.\frac{\mathrm{d}\mathrm{Im}[r_{s\rightarrow s}(\Phi^\prime)]}{\mathrm{d}\Phi^\prime}\right|_{\Phi^\prime=0}.
\end{align}
$\beta_\mathrm{m}$ is the modulation index of the EOM, $P_\mathrm{GR}$ is the GR laser power. The DC power can be written as:
\begin{align}
    P_\mathrm{DC}^\mathrm{DRC} = |r_{s\rightarrow s}(0)|^2 P_\mathrm{GR}. \label{eq:P_DC_DRC}
\end{align}
Substituting Eq. (\ref{eq:D_DRC}) and (\ref{eq:P_DC_DRC}), for (\ref{eq:length_shot_noise}), one can obtain the shot noise level of the DRC:
\begin{align}
    S^\mathrm{DRC} &= \sqrt{2e\frac{e\eta P_\mathrm{DC}^\mathrm{DRC}}{2\hbar\omega_0}}/D^\mathrm{DRC}.
\end{align}
\indent Also in the case of the dithering control, one can use the same approach as~\cite{2001AmJPh..69...79B}. 
The slope amplitude $D^\mathrm{Dither}$ is:
\begin{align}
    D^\mathrm{Dither}\delta l_\mathrm{PCC} &\sim 2J_1(\beta)P_c\mathrm{Im}[1-e^{-i\delta\phi_\mathrm{PCC}}] \label{eq5}\\
    &= \frac{8\pi\beta}{\lambda_0}P_c \delta l_\mathrm{PCC}, \label{eq6}
\end{align}
where $P_c$ is the carrier power leaking from the BS to the QWP by the DC offset, $A$ is the amplitude of the PCM modulation, $\lambda_0$ is the wavelength of the main laser, $J_n$ is the $n$-th order Bessel functions and $\beta$ is the modulation index. For the transformation from Eq.~(\ref{eq5}) to Eq.~(\ref{eq6}) we have used
\begin{align}
    J_1(\beta) &= \frac{\beta}{2} \equiv \frac{2\pi A}{\lambda_0}, \notag\\
    \delta\phi_\mathrm{PCC} &= \frac{4\pi\delta l_\mathrm{PCC}}{\lambda_0}. \notag
\end{align}
$P_\mathrm{DC}^\mathrm{Dither}$ can be written as:
\begin{align}
    P_\mathrm{DC}^\mathrm{Dither} = |1-F(\psi_0)|^2 P_c, \label{eq:P_DC_Dither}
\end{align}
where $F$ is the arm cavity reflectivity:
\begin{align}
   F(\psi_0) = -r_1+\frac{t_1^2 r_1 e^{-i\psi_0}}{1-r_1 r_2 e^{-i\psi_0}}
\end{align}
and $\psi_0$ is the round-trip phase shift of the arm cavity for the IR. Substituting Eq. (\ref{eq6}) and (\ref{eq:P_DC_Dither}) for (\ref{eq:length_shot_noise}), one can obtain the shot noise level of the dithering control:
\begin{align}
    S^\mathrm{Dither} &= \sqrt{2e\frac{e\eta P_\mathrm{DC}^\mathrm{Dither}}{\hbar\omega_0}}/D^\mathrm{Dither}.
\end{align}
The ratio of the two control methods can be written as:
\begin{align}
    \frac{S_L^{\mathrm{Dither}}}{S_L^\mathrm{DRC}} &= \frac{4\beta_m}{\beta}\xi\left|\frac{1-F(\psi_0)}{r_{s\rightarrow s}(0)}\right|\sqrt{\frac{P_\mathrm{GR}}{P_c}}. \label{eq:ratio}
\end{align}

\begin{table*}[t]
\caption{\label{tab:loss}
\textbf{Losses and errors used for the simulation.}}
\begin{ruledtabular}
\begin{tabular}{ccc}
    Parameters & value & Note  \\ \hline
    $\mathcal{L}_\mathrm{arm}$ &  130 ppm  & Arm cavity round-trip loss including ETM transmissivity, $30$ ppm  \\
    $\mathcal{L}_\mathrm{BS, QWP, PBS, PCM}$ & 50 ppm  & PCC optics losses \\ 
    $T_\mathrm{SPBS}$ & 1 \% & PBS $s$-pol transmissivity \\ 
    $R_\mathrm{PPBS}$ & 1 \% & PBS $p$-pol reflectivity \\ 
    $\mathcal{L}_\mathrm{align}$ & 1 \% & Loss by PCM misalignment \\ 
    $\mathcal{L}_\mathrm{mis}$ & 0.15 \% & Mode mismatch between X and Y arm \\ 
    $\delta\phi_{\mathrm{PCC}}$ [rad] & $10^{-7}$ & PCC round-trip phase fluctuation by the PLL noise \\
\end{tabular}
\end{ruledtabular}
\end{table*}

\bibliography{cite}
\end{document}